**Improving the Utilization of Digital Services -** Evaluating Contest-Driven Open Data Development and the Adoption of Cloud Services

Workneh Yilma Ayele
DSV Report Series
Series No. 18-008

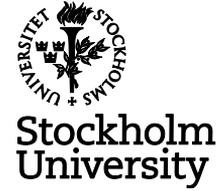

# Improving the Utilization of Digital Services

Evaluating Contest-Driven Open Data Development and the Adoption of Cloud Services

Workneh Yilma Ayele

# Abstract

There is a growing interest in the utilization of digital services, such as software apps and cloud-based software services. The utilization of digital services enabled by ICT is increasing more rapidly than any other segment of the world trade. The availability of open data unlocks the possibility of generating huge market possibilities in the public and private sectors such as manufacturing, transportation, and trade. Digital service utilization can be improved by the adoption of cloud-based software services and through open data innovation for service development.

However, open data has no value unless utilized and little is known about the development of digital services using open data. The use of contests to create awareness and call for crowd participation is vital to attract participation for digital service development. Also, digital innovation contests stimulate open data service development and are common means to generate digital services based on open data. Evaluation of digital service development processes stimulated by contests all the way to service deployment is indispensable. In spite of this, existing evaluation models are not specifically designed to measure open data innovation contest.

Additionally, existing cloud-based digital service implications, opportunities and challenges, in literature are not prioritized and hence are not usable directly for adoption of cloud-based digital services. Furthermore, empirical research on user implications of cloud-based digital services is missing.

Therefore, the purpose of this thesis is to facilitate the utilization of digital services by the adoption of cloud-based digital services and the development of digital services using open data. The main research question addressed in this thesis is: "*How can contest-driven innovation of open data digital services be evaluated and the adoption of digital services be supported to improve the utilization of digital services?*"

The research approaches used are design science research, descriptive statistics, and case study for confirming the validity of the artifacts developed. The design science approach was used to design new artifacts for evaluating open data service development stimulated by contests. The descriptive statistics was applied on two surveys. The first one is for evaluating the implication of cloud-based digital service adoption. While the second one is a longitudinal survey to measure perceived barriers by external open data digital service developers.

In this thesis, an evaluation model for digital innovation contest to stimulate service development, (Digital Innovation Contest Measurement Model) DICM-model, and (Designing and Refining DICM) DRD-method for designing and refining DICM-model to provide more agility are proposed. Additionally, the framework of barriers, constraining external developers of open data service, is also presented to better manage service deployment to enable viable service development. Organizers of open data innovation contests and project managers of digital service development are the beneficiaries of these artifacts. The DICM-model and the DRD-method are used for the evaluation of contest and post contest deployment processes. Finally, the framework of adoption of cloud-based digital services is presented. This framework enables requirement engineers and cloud-based digital service adoption personnel to be able to prioritize factors responsible for an effective adoption. The automation of ideation, which is a key process of digital service development using open data, developer platforms assessment to suggest ways of including evaluation of innovation, ex-post evaluation of the proposed artifacts, and the expansion of cloud-based digital service adoption from the perspectives of suppliers are left for further investigations.

# List of publications

This thesis comprises of the following publications:

I. <u>Workneh Y. Ayele,</u> Gustaf Juell-Skielse, Anders Hjalmarsson and Paul Johannesson. "Evaluating Open Data Innovation: A Measurement Model for Digital Innovation Contests." *PACIS 2015 proceedings* (2015).

II. <u>Workneh Y. Ayele,</u> Gustaf Juell-Skielse, Anders Hjalmarsson and Paul Johannesson (2016): A Method for Designing Digital Innovation Contest Measurement Models: http://media.sigprag.net/2016/11/AyeleEtal-PractDID2016.pdf

III. Anders Hjalmarsson, Gustaf Juell-Skielse, <u>Workneh Y. Ayele</u>, Daniel Rudmark, & Paul Johannesson (2015). From Contest to Market Entry: A Longitudinal Survey of Innovation Barriers Constraining Open Data Service Development. In *ECIS 2015*.

IV. <u>Workneh Y. Ayele</u>, and Gustaf Juell-Skielse. "User Implications for Cloud Based Public Information Systems: A Survey of Swedish Municipalities." *Proceedings of the 2015 2nd International Conference on Electronic Governance and Open Society: Challenges in Eurasia*. ACM, 2015.

# Contents

# Abbreviations

| Abbreviations | Meaning |
|---|---|
| BSc | Balanced Scorecard |
| DICM-model | Digital Innovation Contest Measurement Model |
| DICo | Digital Innovation Contest |
| DRD-method | Designing and Refining Digital Innovation Contest Measurement Model |
| DSR | Design Science Research |
| GQM | Goal Question Metrics |
| IoT | Internet of Things |
| IVC | Innovation Value Chain |
| PDCA | Plan Do Check Act |
| QIP | Quality Improvement Paradigm |
| RQ | Research Question |
| SaaS | Software as a Service |

# 1. Introduction

In this thesis, the process from open data digital service prototype to digital service development and the adoption of cloud-based services is investigated. The improvement of the utilization of digital services through contest-driven open data development and the adoption of cloud-based services is investigated by:

- the evaluation of the processes involving digital service prototype development and the deployment of prototypes to viable digital services
- evaluating barriers constraining external developers to enter the market and by providing cloud-based digital services adoption framework to improve the implementation.

The utilization of digital services is investigated through the phase of con-test-driven prototype development from ideas to viable digital services. The phase from idea to viable service is researched using evaluation of contest-driven development. The succeeding phase, when viable services are deployed, is researched by evaluating barriers constraining developers of open data services, and through the investigation of adoption of cloud-based services. The four papers introduced in the List of Publications section are the foundation for this thesis.

## Digital Services

Digital services, in the discourse of this thesis and the four included papers, refer to apps or services developed from open data and cloud-based systems provisioned in a SaaS service model. However, digital services can mean a wide variety of services provisioned through the internet and operated on multiple platforms. Digital services can be software apps running on mobile devices (Ghazawneh 2016), an arrangement of digital transactions including information, software modules, or consumer goods over the internet (William et al. 2008). A wide variety of services provisioned on the web is considered digital services. For example, e-Commerce services, social media websites, travel management apps and services, online gaming websites, mobile app stores, cloud-based software services, such as, Salesforce.com CRM (William et al. 2008). Rizk et al. present a cloud-based SaaS service model as a digital infrastructure and as a pillar of digital service innovation by summarizing literature about big data analytics and digital service innovation. This cloud-based digital infrastructure delivers digital services with a SaaS service model such as Data-as-a-Service, Information-as-a-Service, and Analytics-as-a-Service (Rizk et al. 2017). There is an increasing demand for these digital services acclaimed by the economic potential in societal and business values.

The growing interest in the utilization of digital services is found in all domains and across a variety of sectors in the world economy. For example, digital service utilization enabled by ICT is increasing more rapidly than any other segment of the world trade (Pajarinen et al. 2013) and public cloud service spending worldwide is estimated to reach $195 billion by 2020 (IDC, 2016). There is a shift from a product-based economy to a service, particularly a digital service, based economy, (Williams et al. 2008). Also, the dividing line between services and manufacturing is becoming blurred both at the aggregate level as well as within companies, for example, IBM has shifted from a manufacturing company to a leading IT service provider (Pajarinen et al. 2013). Also, the nature of service and product innovations in organizations is profoundly being affected by the widely spreading pervasive digital technologies (Yoo et al. 2012). As an illustration, IoT involves the emerging utilization of digital artifacts that combines well-established products with digital solutions (Atzori et al. 2010). The utilization of digital services is, therefore, considered a viable domain of study (Ostrom et al. 2015). This thesis is focused on improving the utilization of digital services.

With the innovation of web-based services and the advanced infrastructures supporting real-time data obtained from sensors and data accumulation, digital services are becoming inevitable in supporting day-to-day activities. Digital services can provide traffic information, climate information, government services, and electronic banking. Digital services can be developed from open data and by the adoption of cloud-based digital services. Freely available data, which can be modified and distributed for any purpose is referred as open data (http://opendefinition.org/). Open data, itself can be delivered as Open-Data-as-a-Service (Segura et al. 2014).

Open data shared by governments enables public-private partnership by instigating innovation and startups and improves services provision (Hendler et al. 2012). Open data shared by governments may cover different sectors such as health (Hendler et al. 2012), transportation (Chui et al. 2014), GIS (Goodchild et al. 2012). Besides, open data can be used in the private sector, for example in manufacturing (Wu et al. 2016). Segura et al. propose an architecture which delivers open data in the form of services, called Open-Data-as-a-Service, for the development of open data applications (Segura et al. 2014). Open data with relevant API information accelerates the development of digital services (Chan 2013). Open data can be delivered along with terms and conditions of use, API information to create an open innovation platform (Chan 2013).

Granting public access to open data facilitates the creation of societal and business values (Lindman et al. 2013).

Open data is becoming increasingly popular means for service development and brings several values to the society. For example, open data unlocks the possibility of advancing entrepreneurship, stimulate start-ups, and enhance services (Lakomaa and Kallberg 2013). Also, cloud-based digital services enable the realization of efficiency gains as well as innovation and entrepreneurship (Kushida et al. 2011). Digital services provisioned to citizens increases the convenience and accessibility of government services and information (Carter and Bélanger 2005). Therefore, the utilization of digital services can be facilitated by the use of open data and cloud-based digital services.

Digital service utilization can be improved, as discussed in this thesis:

- by the development of viable digital services using open data.
- by managing barriers constraining developers from developing viable services and by supporting the adoption of cloud-based digital services.

The development of digital services can be improved by the provision of open data. The growth of open data regarding size is exponential (Kundra 2012). Its direct market, only in EU, is projected to grow by 36.9% between 2016 and 2020 (Carrara et al., 2015). Also, there are plenty of open data markets for the public to advance entrepreneurship, stimulate start-ups, and enhance services (Lakomaa and Kallberg 2013). Open data is estimated to generate 3$ to 5$ trillion dollars annually across different sectors such as education, transportation, consumer products, electricity, oil and gas, health care, and consumer finance worldwide (Chui et al. 2014). Open data has the potential to increase the availability of most governmental datasets for service development (Lindman et al. 2013). Goodchild et al. argue that in the geographic information arena the future will witness a significant progress in the utilization of open data gaining momentum in several countries (Goodchild et al. 2012). Both the public and private sectors can publish and use open data.

The adoption of cloud-based digital services in the private sector, specifically Cloud ERP, empowers users and suppliers with several opportunities (Juell-Skielse and Enquist 2012). Also, the utilization of digital services can be improved by the adoption of cloud-based digital services in the private sector. For example, in the manufacturing sector, cloud-based digital services transform the life cycle of product development from design, analysis, manufacturing, and to production (Wu et al. 2016). Juell-Skielse and Enquist (2012) developed a framework of implications, opportunities and challenges, of adopting cloud-based services from the perspectives of users and suppliers of SaaS.

# Research Challenge

Evaluation of innovation contest and barriers constraining viable digital services development and the adoption of cloud-based digital services are the main areas of this research. Organizers of contests, users of digital services, and developers are the main stakeholders for the artifacts presented in this research. Organizers of contest-driven development of digital services need to evaluate their digital service development and deployment processes. Developers also need to be able to cope with barriers in post-contest processes to develop viable digital services. Also, users demand to use sustainable digital services. Besides, users and developers require adopting cloud-based services that meet their requirements efficiently.

Open data has no value if not utilized and the development of digital services using open data is in its infancy (Janssen et al. 2012). The use of contests to create awareness and call for crowd participation is vital for attracting participation in digital service innovation strategy (Chan 2013). Digital innovation contests enable open data knowledge sharing (Smith et al. 2016), stimulates open data service development (Juell-Skielse et al. 2014), and are common means to generate digital services based on open data (Hjalmarsson et al. 2014). Effective evaluation of contests, therefore, is an essential endeavor in managing the contest-driven development of digital services.

Also, to support and evaluate open data service development and the effects of digital innovation contests, the Innovation Value Chain of Hansen and Birkinshaw (2007) for open digital services need to be better managed. Furthermore, only a limited number of open data services developed by contests reach the market due to innovation barriers faced by external developers (Hjalmarsson et al. 2014).

Research regarding constraining barriers by (Lüttgens et al. 2014; Ghobadi and Mathiassen 2014) using primarily qualitative research approaches has provided sound insights but the impact of the barriers over time still needs research attention. Hence, evaluation of digital service development processes stimulated by contest all the way to service deployment and the investigation of barriers constraining external developers is indispensable. Finally, existing list of implications, opportunities and challenges, of adoption of cloud-based services are not prioritized to enable effective adoption (Juell-Skielse and Enquist 2011). There are numerous user implications identified by several authors (Juell-Skielse and Enquist 2011) which can be further investigated.

## Research Purpose

The purposes of the artifacts presented in this thesis are to facilitate the utilization of digital services by contest organizer, users, and developers. The research question and its two sub-questions are presented below:

How can contest-driven innovation of open data digital services be evaluated and the challenges constraining the development and the adoption of cloud-based digital services be supported to improve the utilization of digital services?

1. How can the process of contest-driven digital innovation be evaluated?
2. What are the barriers constraining the development of digital services, and the challenges to effectively to adopt cloud-based digital services?

The answers to these questions are covered in the four papers and the artifacts developed are illustrated in Table 1.

| RQs | Artefacts |
|---|---|
| RQ. 1. | • Model for measuring digital innovation contests, DICM-model (Paper I)<br>• Method for designing DICM-models, DRD-method (Paper II) |
| RQ. 2. | • Framework of innovation barriers constraining open data service development in different phases by external developers (Paper III)<br>• Framework for adopting cloud-based digital services (Paper IV) |

Table 1. Research questions and the artefacts developed

This thesis is organized into six chapters. This chapter presents the introduction, and the next chapter presents the theoretical grounding. The methodology, results, discussion and future research, and conclusion are presented in the succeeding chapters.

## Brief Summary of Publications

Paper I

| Title | Evaluating Open Data Innovation: A Measurement Model for Digital Innovation Contests |
|---|---|
| Submitted to | The Pacific Asia Conference on Information Systems (PACIS) |
| Publisher | Association for Information Systems, http://home.aisnet.org/associations/7499/files/Index_Markup.cfm, USA |

| Research questions | Are existing innovation measurement models applicable for evaluating open data innovation catalyzed by innovation contest and what are the components of DICMs? |
|---|---|
| Research Methods | Design Science Research and using a single case study |
| Premises for the research | Viktoria Swedish ICT, University of Borås and Stockholm University |
| Contribution | A summary of innovation measurement models, a measurement model for pre and post digital innovation contest processes |
| Responsibility | Workneh Y. Ayele wrote 50 % of the article. |

## Paper II

| Title | A Method for Designing Digital Innovation Contest Measurement Models |
|---|---|
| Submitted to | AIS SIGPRAG Pre-ICIS Workshop 2016 "Practice-based Design and Innovation of Digital Artifacts" |
| Publisher | Association for Information Systems, http://home.aisnet.org/associations/7499/files/Index_Markup.cfm, USA |
| Research questions | What are the problems related with applicability of DICM model and how can a suitable solution for designing applicable DICMs be designed? |
| Research Methods | Design Science Research |
| Premises for the research | Viktoria Swedish ICT, University of Borås and Stockholm University |
| Contribution | To support open data development by evaluating the process of development of digital artifacts stimulated by contests |
| Responsibility | Workneh Y. Ayele wrote 75 % of the article. |

## Paper III

| Title | From contest to market entry: a longitudinal survey of innovation barriers constraining open data service development |
|---|---|
| Submitted to | ECIS 2015 - European Conference on Information Systems |
| Publisher | Association for Information Systems, http://home.aisnet.org/associations/7499/files/Index_Markup.cfm, USA |
| Research questions | What innovation barriers constrain external developers in different phases when performing open data service development after innovation contests? |
| Research Methods | Quantitative, longitudinal survey<br>Single case study |
| Premises for the research | Viktoria Swedish ICT, University of Borås and Stockholm University |

| Contribution | To support open data development by identifying barriers perceived by external developers in deferent phases of the development of digital artifacts after innovation contests |
|---|---|
| Responsibility | Workneh Y. Ayele wrote 25 % of the article. |

## Paper IV

| Title | User Implications for Cloud Based Public Information Systems: A Survey of Swedish Municipalities |
|---|---|
| Submitted to | Electronic Governance and Open Society: Challenges in Eurasia (EGOSE 2016) |
| Publisher | Association for Computing Machinery (ACM), http://www.acm.org/pubs/, USA |
| Research questions | What are the most significant implications for adopting cloud-based digital service? |
| Research Methods | Quantitative |
| Premises for the research | Stockholm University, DSV |
| Contribution | Adoption of cloud-based public information systems in particular ERP systems |
| Responsibility | Workneh Y. Ayele wrote 80 % of the article. |

# 2. Theoretical Background

In this chapter, an overview of concepts covered in this thesis, see Figure 1 and theoretical foundations are presented. The innovation of open data through contest-driven development and the adoption of digital services to enable the utilization of digital services are the primary concepts in this thesis.

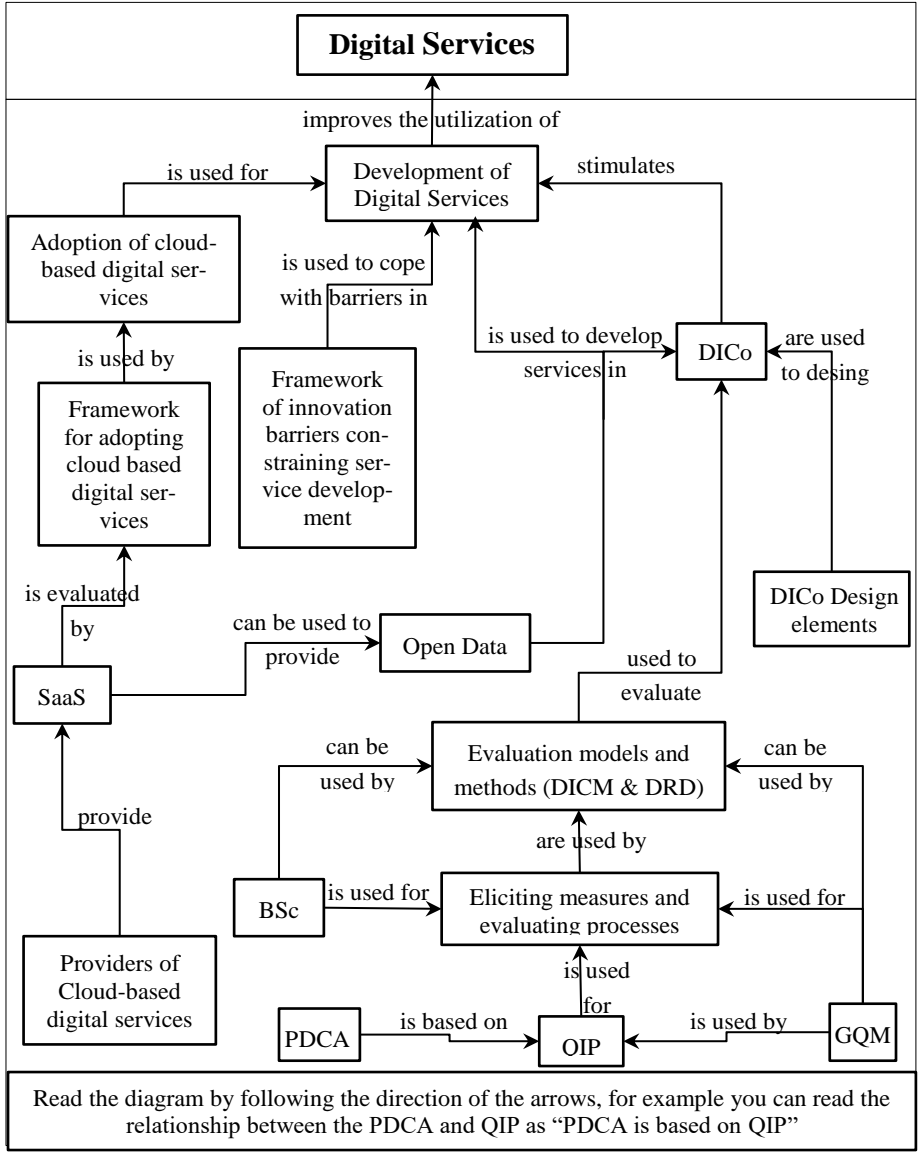

*Figure 1: Central concepts, illustrating how digital innovation contests and adoptions of cloud-based digital services lead to the facilitation of the utilization of digital services.*

# Digital Services

There are distinctions between services and digital services. Williams et al. 2008 described digital services as the arrangement of digital transactions such as information, software modules, or consumer goods over the internet, while service is an activity that one party can give to another (William et al. 2008). Digital services can be developed by utilizing open data (Hjalmarsson et al. 2014). Segura et al. propose an architecture which delivers open data on the cloud, Open-Data-as-a-Service, for the development of open data applications (Segura et al. 2014).

Also, digital services can have a variety of translation in the realm of the digitization. For example, mobile apps (Ghazawneh 2016), web-based electronic commerce applications (William et al. 2008), cloud-based CRM such as Salesforce (William et al. 2008, p. 508), Cloud-based services such as Analytics-as-a-Service (Rizk et al. 2017) are all considered as digital services. In this thesis, digital services may mean mobile apps, cloud-based services offered by any sector, web-based services, and GIS services. All these technologies improve the utilization of digital services.

# Adoption of Cloud-Based Digital Services

Public organizations can gain scalability and faster processing by deploying cloud-based digital services (Armbrust et al. 2010). Digital services can be assembled by subscribing cloud services from vendors in a pay per use model (Abels et al. 2010). The adoption of cloud-based digital services improves the utilization of digital services in different sectors. For example, health care (Kuo 2011; Nkosi and Mekuria 2010), E-Government (Lian 2015), manufacturing (Xu 2012), and education (Behrend et al. 2011; Sabi et al. 2016).

## Cloud Computing

"Cloud computing is a model for enabling ubiquitous, convenient, on-demand network access to a shared pool of configurable computing resources (e.g., networks, servers, storage, applications, and services) that can be rapidly provisioned and released with minimal management effort or service provider interaction. This cloud model is composed of five essential characteristics, three service models, and four deployment models." (Mell and Grance, 2011). The essential characteristics are on-demand self-service, broad network access, resource pooling to facilitate accessibility by the multitude, rapid elasticity to facilitate scalability and measured services (Mell and Grance, 2011).

Cloud has four deployment models classified according to ownership, operation, and management such as private cloud, community cloud, public cloud, and hybrid cloud (Mell and Grance, 2011).

The three service models of cloud computing are discussed in this paragraph. The first one is Software-as-a-Service (SaaS), where applications can be accessed from client devices and applications are provided to consumers by service providers. The second one is Platform-as-a-Service (PaaS), where consumers can deploy applications that they created or acquired and the providers avail cloud infrastructure, libraries, services and programming languages. And the last one is Infrastructure-as-a-Service (IaaS), where consumers can deploy applications and operating systems on the infrastructure providers avail such as processing, storage, networks, and other fundamental computing resources (Mell and Grance, 2011). Digital services can be deployed on the SaaS model such as Open-Data-as-a-Service (Segura et al. 2014); Data-as-a-Service, Information-as-a-Service, and Analytics-as-a-Service (Rizk et al. 2017); ERP-as-a-Service (Juell-Skielse and Enquist 2012).

## Framework for Adopting Cloud-Based Digital Services

Implications are the consequences of adopting cloud computing. Some authors use different terms to refer to implications for example opportunities and challenges (Cusumano 2008), and benefits and risks (Sääksjärvi et al. 2005). Juell-Skielse and Enquist developed a framework of implications, consisting of more than 80 implications, for adopting cloud computing by categorizing implications according to "opportunities" and "challenges" and divide between user and supplier implications (Juell-Skielse and Enquist 2012). Implications of cloud computing can be used for the adoption of cloud-based services (Kuo 2011). However, the framework of implications by Juell-Skielse and Enquist is not organized by assessing the significance of each implication and presenting them regarding the order of significance to aid the adoptions in a logical manner (Juell-Skielse and Enquist 2012).

# Development of Digital Services

Development of digital services increases the availability of digital services hence the utilization is improved. Digital services can be developed using open data stimulated contests, DICos (Hjalmarsson et al. 2014). External developers use open data for service development driven by commercial (Ceccagnoli et al. 2012) or non-commercial motives (Kuk and Davies 2011).

## Digital Innovation Contest

Digital innovation is the process in which a product, process, or business model that is regarded as new, or that involves some significant changes, is developed and is embodied in or enabled by IT (Fichman et al. 2014). Hjalmarsson and Rudmark define <u>digital innovation contest as the contest among external developers to design and implement the most secure and satisfying digital service prototype based on open data for a specific purpose (Hjalmarsson and Rudmark 2012, p. 10). Contests have become popular</u> to stimulate digital innovation to generate digital services based on open data (Hjalmarsson et al. 2014). Also, contests are used to create awareness and attract participation in digital service development (Chan 2013). Design elements of digital innovation contests are used to design key elements of innovation contests and the post-contest processes (Juell-Skielse et al. 2014).

## Open Data

Open data is defined as: *"Open data and content can be freely used, modified, and shared by anyone for any purpose",* Open Definition[1]. However, some people understand open data as data available internally in an organization while others understand it as data that is available for everyone externally (Tammisto and Lindman 2012). There is an increasing demand for opening data provided by the public and private organizations (Zuiderwijk et al. 2012). Besides large software vendors are also opening up their data to their cloud platforms such as Microsoft's Window Azure (Kirchhoff and Geihs 2013).

In this thesis, open data is regarded as data availed to external developers and the community to stimulate innovation, often accessible through API in a machine readable manner and labeled as open architecture (Marton et al. 2013). In literature, open data is defined by Open Definition.

## Barriers Constraining Digital Service Developers

In this thesis, barriers are innovation barriers that developers face to transform prototypes into viable digital services over time after contest process. Hjalmarsson et al. designed a framework of barriers, constraining external developer from developing viable digital services (Hjalmarsson et al. 2014).

---

[1] Open Definition (N. D). "The Open Definition". http://opendefinition.org/

## Evaluation models

There are various models and frameworks available for evaluating innovation. However, there is hardly any model designed to assess open data digital innovation contest processes. Available evaluation models can be used for different purposes. For example, evaluation of openness regarding open innovation (Enkel et al. 2011), evaluation of product-, process-, and organizational & marketing- innovation (OECD 2005). Finally, evaluation of innovation at the country, industry and organizational level (Mairesse and Mohnen 2002), and the contribution and effectiveness of an organization in an innovation ecosystem (Porter 1990).

The evaluation models available in the literature have characteristics which dictates how the evaluation is done. For example, some evaluation modes are based on the Innovation Value Chain (IVC) which is characterized to identifying weakness and strengths in the IVC (Hansen and Birkinshaw 2007; Ress et al. 2013). There are also models which adopt the BSc of Kaplan and Norton 1996 (Erkens et al. 2013; Flores et al. 2009). Finally, innovation evaluation models can also be based on goal-oriented approaches such as GQM (Misra et al. 2005).

## Eliciting Measures and Evaluating Processes

Evaluating models can use techniques such as GQM and or BSc to elicit measures. Evaluation of processes can implement the PDCA and IVC for process improvements. There are innovation measurement models or frameworks that use goals and questions to identify metrics. The model by Misra et al. (2015) employs goals to define measures of innovation using GQM (Misra et al. 2005). However, Misra et al. did not include a clear way in which goal, question, and metric are defined (Edison et al. 2013). In the contrary Hansen and Birkinshaw used key questions to identify measures (Hansen and Birkinshaw 2007). Results of questions can be used to measure innovation, Enkel et al. used answers to questions to measure maturity levels (Enkel et al. 2011). Likewise, a set of questions is asked to assess dimensions of innovation in Diamond Model (Tidd et al. 2002; Gamal et al. 2011). Therefore, a GQM paradigm is deliberately or otherwise applied in identifying measures in some models. Finally, since higher level strategic perspective is missing in GQM, strategic goals can be addressed using BSc (Buglione and Abran 2000).

## Goal Question Metric

The GQM was formerly proposed for evaluation of defects in software engineering projects in NASA (Basili and Weiss 1984) and is commonly used in software projects (Buglione and Abran 2000). However, GQM is also relevant in other disciplines such as software and information security (Savola 2008; Kowalski and Barabanov 2011; Kassou and Kjiri 2013), information systems (Esteves et al. 2003; Ganesan and Paturi 2009; Kassou and Kjiri 2013), and healthcare (Villar 2011). In the GQM paradigm, questions are derived from goals and metrics are derived from questions (Basili et al. 1994).

The GQM has three levels: conceptual level (Goal) - measurement goals are defined for a product, processes or resources. Operational level (Question) - set of question are formulated to meet the specified goals, and finally, quantitative level (Metric) – answers to questions are identified. The measured data can be objective such as worker's hours spent on task and can also be subjective for example level of satisfaction (van Solingen et al., 2002). Measures can be turned to quantitative values using scales presented by (Hansen and Birkinshaw 2007; Erkens et al. 2013).

## Balanced Scorecard

The BSc is a strategic means that facilitates identify strategic measures to assess the impact of open innovation (Flores et al. 2009). BSc is a multidimensional framework that relates objectives, initiatives, and measures to an organization's strategy at all levels (Kaplan and Norton, 1996). The GQM and the BSc cover different aspects of measures, consequently the use of both approaches are fruitful (Buglione and Abran 2000). The BSc and the GQM have similarities in defining metrics. For example, BSc uses goal-driver-indicator to define metrics in a similar fashion as goal-question-metrics does. The difference is that the GQM is relevant to multiple contexts while the BSc has a structure to facilitate the alignment of operational goals and business goals (Buglione and Abran 2000).

## Plan Do Check Act

The Shewart-Deming Cycle Plan-Do-Check-Act commonly known as PDCA is a widely known model for continuous process improvements, it explicates how organizations plan, do the planned task, check to make sure the planned task is actually done and act on what has been learned (Johnson 2002).

## Quality Improvement Paradigm

The QIP has six steps cycle emphasizing continuous improvement and is based on PDCA (Basili et al. 1994). QIP uses the GQM paradigm for evaluating and articulating a list of operational goals (Basili et al. 1994). The six steps of QIP are 1) characterize: understand environment and establish baseline with existing business process also use knowledge gained from previous projects, 2) set goals: based on 1) identify goals that lead to success in the project, 3) choose process: based on 1) and 2) choose suitable processes, 4) execute: execute processes and provide project feedback based on data collected on goal achievement, 5) analyze: at the conclusion of the project collect data and make analysis to assess current practice, identify problems, and make recommendations for future projects, and 6) package: combine knowledge gained from the current project with previous projects and store it for future projects.

# Utilization of Digital Services

According to Oxford online dictionary, the word utilize originates in the early 19th century from Italian utilizzare and from French utiliser, utilize is "make practical and effective use of" whereas use is "take, hold, or deploy (something) as means of accomplishing or achieving something; employ". Since improving utilization of digital services entails effective, and practical use of digital services, in this thesis, digital service utilization means – an effective and practical use of digital services.

The utilization of digital services enabled by ICT (Pajarinen et al. 2013) enhances the convenience and accessibility of public services (Carter and Bélanger 2005). The utilization also empowers users and suppliers of cloud-based services with substantial opportunities (Juell-Skielse & Enquist 2012).

Contest-driven open-data service development and the adoption of cloud services improve the utilization of digital services. The use of open data with relevant API information provided accelerates the development of digital services (Chan 2013). Open data is growing exponentially (Kundra 2012). Also, it is estimated to generate trillions of dollars annually across different sectors worldwide (Chui et al. 2014). However, open data has no value unless utilized and little is known about utilizing open data to develop digital service (Janssen et al. 2012).

Utilization of digital services can be improved by open innovation contest to stimulate the use of open data for service development and by the adoption of cloud-based services. Open innovation includes the creation of open innovation platforms and attracting the participation of potential partners (Chesbrough and Appleyard 2007). The provision of relevant APIs along with the datasets in the open innovation platform accelerates the development of digital services (Chan 2013). The adoption of the cloud in a variety of sectors such as education (Behrend et al. 2011; Sabi et al. 2016), health care (Kuo 2011; Nkosi and Mekuria 2010), manufacturing (Xu 2012), and E-Government (Lian 2015).

Contests are popular to stimulate digital innovation to develop digital services based on open data (Hjalmarsson et al. 2014), and create awareness and attract participants (Chan 2013). However, the development of digital services using open data is still in its early stages (Janssen et al. 2012). Additionally, during the post-contest process, external developers face barriers constraining them to build viable digital services from prototypes (Hjalmarsson et al. 2014).

The adoption of cloud-based digital services has transformed different sectors such as manufacturing (Wu el al. 2016; Xu 2012), education (Behrend et al. 2011; Sabi et al. 2016), health care (Kuo 2011; Nkosi and Mekuria 2010), and E-Government (Lian 2015). Additionally, cloud-based digital services enable the realization of innovation and entrepreneurship (Kushida et al. 2011).

# 3. Methodology

The research objective of this thesis is to investigate how the utilization of digital services be improved by answering the research question and its sub-questions stated below.

How can contest-driven innovation of open data digital services be evaluated and the challenges constraining the development and the adoption of cloud-based digital services be supported to improve the utilization of digital services?

1. How can the process of contest-driven digital innovation be evaluated?
2. What are the barriers constraining the development of digital services, and the challenges to effectively to adopt cloud-based digital services?

The answers to these questions are addressed by applying the research methods, and the corresponding research contributions are illustrated below in Table 1.

| RQs | Contributions | Stakeholders | Research method and approaches used |
|---|---|---|---|
| RQ. 1. | DICM-model (Paper I) | Organizers | DSR, Literature review, Case study |
| | DRD-method (Paper II) | Organizers | DSR, Literature review |
| RQ. 2. | Framework of innovation barriers constraining open data service (Paper III) | Organizers, Developers | Quantitative method, Case study, … |
| | Framework for adopting cloud-based digital services (Paper IV) | Users, Developers | Quantitative method |

*Table 1: Research questions and methods and approaches uses to address them with corresponding contributions and stakeholders*

The utilization of digital services can be improved through the evaluation of the digital service development stimulated by contests, the analysis of barriers constraining external developers from building viable digital services, and the adoption of cloud-based digital services.

# Philosophical Assumptions

Philosophical assumptions inform the choice of research process and theories. The dimensions of these assumptions are ontological beliefs, epistemology, axiological beliefs, and methodological beliefs. Ontological belief describes the nature of reality while epistemology describes how reality is known. Finally, axiological belief stresses on the role of values while methodological belief emphasizes on the approach to inquiries (Creswell 2013).

In this thesis, two philosophical assumptions frameworks, illustrated in Table 2, are used to explain the underlying philosophical assumption in the four papers. The first philosophical assumptions framework is applicable for qualitative studies as adopted from Lincoln et al. (2011) see the first two columns in Table 2 below. The second philosophical assumptions framework is applicable for DSR artifact development presented by Vaishnavi, and Kuechler (2015) see the last two columns in Table 2 below. As illustrated in Table 2 below positivism and design paradigms are adopted from (Vaishnavi, and Kuechler 2015) and social constructivism, which is similar to interpretive (Creswell 2013), and pragmatism are adopted from (Lincoln et al. (2011) as cited by Creswell 2013, p. 36-37). The underlying philosophical assumptions illustrated in Table 2 below are applied in the four papers see as described in Table 3 below.

| Paradigms | Ontology | Epistemology | Axiology | Methodology |
|---|---|---|---|---|
| Interpretative (social constructivism) | Multiple realities are constructed through our lived experiences and interaction with others. | Reality is co-constructed between the researcher and the researched and shaped by individual experiences. | Individual values are honored, and are negotiated among individuals. | More of a literary style of writing used. Use of an inductive method of emergent ideas (through consensus) obtained through methods such as interviewing, observing, and analysis of text. |
| | (Lincoln et al. 2011 as cited in Creswell 2013, p. 36-37) | | | |
| Pragmatism | Reality is what is useful, is practical, and "worked" | Reality is known through using many tools of research that reflects both deductive (objective) evidence and inductive (subjective) evidence | Values are discussed because of the way that knowledge reflects both the researchers' and the participants' views. | The research process involves both quantitative and qualitative approaches to data collection and analysis. |
| | (Lincoln et al. 2011 as cited in Creswell 2013, p. 36-37) | | | |
| Positivism | A single reality, knowable, probabilistic. Reality is external to the researcher and represented by objects in space (Mack 2010). | Objective, dispassionate, Detached observer of truth. *Knowledge is deductively generated (Mack 2010)* | Truth: universal and beautiful; prediction. | Observation; quantitative, statistical. |
| | (Vaishnavi and Kuechler 2015, p. 31) | | | |
| Design | Multiple, contextually situated alternative world-states. Sociotechnologically enabled | Knowing through making: objectively constrained construction within a context. Iterative circumscription reveals meaning | Control; creation; progress (i.e. improvement); understanding | Developmental. Measure artifactual impacts on the composite system |
| | (Vaishnavi and Kuechler 2015, p. 31) | | | |

*Table 2: Interpretative framework for philosophical assumptions*

A mixed method approach is used in this research. The research methods and approaches applied in the four papers discussed in this thesis are design science research in paper I and II, and quantitative methods specifically descriptive statistics in Paper III and Paper IV. A single case study, Travelhack 2013, was used in Paper I and Paper III as a strategy for confirming the validity of the artifacts developed. The case study in Paper I was used to demonstrate and identify key components of the artefact developed, DICM-model. On the other hand, respondents, external developers, from this case study who decided to develop viable digital services were surveyed to measure perceived barriers over a longitudinal period. The research paradigm and data collection and analysis are illustrated in Table 3 below.

| Papers | Philosophical assumptions | Research Approaches and methods | Data Collection and Analysis |
|---|---|---|---|
| Paper I | Design | Case study, Literature review, DSR | Observation and analysis of case study |
| Paper II | Pragmatism, Design | Literature review, DSR | Interview, enquires, deductive thematic analysis |
| Paper III | Positivist, Interpretive | Quantitative method, Case study, | Longitudinal survey, Descriptive statistics |
| Paper IV | Positivist, Interpretive | Quantitative method | Survey, Descriptive statistical, questionnaire |

*Table 3: Philosophical assumptions, selected methods, and data collection & analysis.*

# Research Approaches

Research methods fall into two categories namely qualitative research and quantitative research (Taylor, 2005). The choice of qualitative or quantitative method may depend on the purpose of the study. For example, Punch argues quantitative research is more often conducted for theory verification while qualitative research is more often conducted for theory generation but not necessarily (Punch, 2013). In Paper III quantitative method is used for theory verification while in Paper I it is used for theory generation. Both approaches, quantitative and qualitative, require scientific procedures called research methods, plan of action regarding data collection, sampling, and data analysis (Gelo, 2012).

## Quantitative Method

Statistical research is broadly categorized into two broad categories, inferential statistics and descriptive statistics (Boslaugh and Watters, 2008). Paper I and Paper III uses descriptive statistics to address the research problems in those papers. In Paper I, exploratory factor analysis is used to identify hidden constraints (Treiblmaier and Filzmoser 2009) and to elicit the underlying constructs (Williams et al. 2010). Similarly, descriptive statistics is used in Paper III using simple arithmetic mean and standard deviation (Balnaves and Caputi 2001), to verify the assumption, that is there exists anticipated variation in perceived barriers through time, *effect size* is used as described by (Field 2013) in longitudinal survey. *Effect size* is a descriptive statistic representing the measurement of the strength of the association among variables under a specific situation (Wilkinson, 1999).

## Case study

The content of a case study depends on its audience, purpose, level, and other factors (Lincoln and Guba 1985). A case study is chosen to get an in depth understanding of situations (Creswell 2013). Therefore, in Paper I the case study is used to verify the validity of the DICM-model and to identify key components, while in Paper III the case study is applied to gain an in depth understanding about barriers perceived by external developers. Case study can be conducted for explanatory, exploratory, and descriptive investigations (Yin 2013).

### Travelhack 2013 – Contest Case Description

Storstockholms Lokaltrafik (SL), Stockholm local traffic in English, is a public transport company serving around 800,000 travelers daily. SL and Samtrafiken, ticketing and journey planning company all over Sweden, founded Trafiklab.se in September 2011. Trafiklab.se provides open data and platform to external developers. In the fall of 2012 SL and Samtrafiken organized Travelhack 2013 to increase the utilization of open data platform, Trafiklab.se and to stimulate the development of novel digital services to make public transport more attractive. The contest hosted 150 participants 20-30 teams remain at the final event. Competitors delivered five novel digital services by developing more attractive public transportation digital services. A year after the completion of the contest, two services were in development, and one service reached top ten downloaded services in the travel category in Sweden.

The contest took place in December 2012 with: *idea*, *preparation*, and *final* phases. It was continued for three months, at the beginning 217 interested participated and registered and during the mid-January 58 ideas were entered targeting any of the three categories listed such as digital services that make public transportation:

- more fun
- more efficient
- more accessible to everyone, especially those with cognitive disabilities.

These ideas were evaluated in mid-February and 25 out of 58 entries were invited for a 24-hour final hackathon in March. The criteria for selecting finalists were innovativeness, market potential, usefulness, and technical feasibility. During the preparation phase, teams were given additional APIs organizations other than Trafiklab.se such as Spotify and Microsoft. The 24-hour hackathon was finalized with 21 teams proving their digital service prototypes and expert jury selected winners.

## Literature review

Identification of relevant literature in an effective way is a core activity for advancing knowledge, and therefore a high-quality review is focused on concepts and is complete to cover relevant literature (Webster et al. 2002). Papers II and III use literature review to identify main design elements of artifacts developed during the study.

## Design Science Research

There are six activities in design science study: problem identification, objectives of a solution, design and development, demonstration, evaluation, and communication (Peffers et al. 2007). Design science research is applicable in different disciplines such as computer science, information science, and engineering (Peffers et al. 2012).

# Data collection, analysis, and research processes

The first research question, RQ. 1., is answered by designing a DICM-model, a DRD-method, and a framework of barriers constraining external developers from developing viable digital services. The DICM-model and the DRD-method are designed using DSR. The framework of digital innovation barriers constraining external developers is evaluated using literature review and longitudinal survey. Finally, second research question, RQ. 2., is answered by developing a framework of implications for adopting cloud-based digital services, the process of evaluating implications of cloud-based digital services is presented in this Section.

A literature study was conducted in Paper I to identify existing innovation evaluation models and interviews and observations were carried out to design and verify the DICM-model. In Paper II, semi-structured interviews were administered to 13 respondents to evaluation the DICM-model. The analysis of the evaluation of the DICM-model was conducted using thematic analysis (Braun and Clarke (2006), deductive thematic analyst was used. For readability and ease of interpretation of the thematic analysis the data was organized into groups following the SWOT labeling to identify strengths, weaknesses opportunities, and threats (Hill and West-brook, 1997). Finally, a survey was administered to collect data about perceived barriers in Paper III and another survey was designed to identify the significance of implications of cloud-based digital services in Paper IV. The analysis of data for Paper III and Paper IV were done using descriptive statistics as presented below.

## Paper I – DICM-model

*Activity 1: Problem identification and motivation*
Innovation is not an easy thing to measure or evaluate. However, some things are easy to measure such as revenue improvement (Malinoski and Perry 2011). Therefore, assessing the effects of innovation contests is a challenge and evaluation of digital innovation contest with the succeeding service deployment processes is significant for managing innovation. The nature of innovation and the absence of digital innovation contest measurement models specifically designed for open data service development are the starting point as challenges of assessing the effects of innovation contests. A case study is a suitable approach for a better understanding of the problem (Creswell 2013), and the case study can be conducted for exploratory investigations (Yin 2013).

*Activity 2: Definition of objectives of the solution*
The objectives of the solution are elicited through a literature study and the analysis of the case study. The key objectives of the model are:
1. measuring fulfillment of the goals of an innovation contest

2. identifying strengths and weaknesses in the IVC by measuring underlying factors affecting the results of DICo
3. supporting organizers in learning and increasing maturity
4. to be easily usable by organizers of contests based on available data.

*Activity 3: Design and development*
The designed measurement model, DICM-model, is inspired by IVC of Hansen and Birkinshaw (2007) and has two processes, the innovation contest, and the service deployment processes. Hansen and Birkinshaw proposed a three phase IVC such as input, process, and output (Hansen and Birkinshaw 2007). These three phases enabled the identification of the three phases in each process of the DICM-model. Innovation contest consists of planning, ideation, and service design; while service deployment consists of preparation, implementation and exploitation. ) The phases in the DICM-model are expressed differently in an expressive manner. Moreover, phases in Erkens et al. model have four components each. These components are input, processes, outputs, outcomes (Erkens et al. 2013). These components of phases are adopted as inputs, activities, outputs and measures in the proposed DICM-model. Finally, the case study enabled the elicitation of key characteristics of the activities.

*Activity 4: Demonstration*
The measurement model was applied to the case study to verify the viability of the DICM-model. The components of the DICM-model are demonstrated using running examples for each of the core parts.

*Activity 5: Evaluation*
The evaluation of the DICM-model, was left for future study and as a continuation, the DRD method was developed based on the evaluation of the DICM-model in Paper II In Paper II an ex-ante evaluation was done using semi-structured interview of 13 experts from five countries such as Sweden nine respondents and Cyprus, Greece, Finland, and Singapore one respondent from each. The collected data was evaluation using a thematic analysis by 13 researchers.

*Activity 6: Communication*
The DICM-model was presented at an international conference, and during the evaluation process, it was communicated to 13 experts.

# Paper II - DRD-method
*Activity 1: Problem identification and motivation*
The evaluation of the DICM-model by 13 experts indicated that the model is acceptable but needs to be adapted to meet requirements of a given contest. Also, the model is not flexible enough to be directly applicable "as is" without adopting it to a given context. There is, therefore, a need for a DRD-method to design and refine a DICM-model in an agile way. This was identified after thematic analysis of interview data.

**Thematic analysis and data collection**
Semi-structured interviews were administered to evaluate the DICM-model, as part of problem identification, involving 13 respondents. Thematic analysis was done on the collected data to identify, analyze, and report themes and patterns (Braun and Clarke (2006). The data was organized by labelling the strengths, weaknesses, opportunities, the SWOT for categorizing the analyzed data to identify strengths, weaknesses, opportunities, and threats from the deductive thematic analysis. The respondents are experts in organizing digital innovation contests besides their academic competence.

*Activity 2: Definition of objectives of the solution*
The rigidity in the DICM-model illustrated in Problem identification and motivation above is attributed to the difference in goals of contests in the respondents' organizations. The objectives of the DRD-method are to aid in designing DICM-models that:
1. measure the fulfillment of the measurement goals of contests
2. identify strengths and weaknesses in the IVC
3. support learning and knowledge management in the development of digital services
4. to be easy for organizers of contests.

*Activity 3: Design and development*
The components of DRD-method were identified by following quality oriented software evaluation QIP paradigm by NASA (Basili et al. 1994). The DRD-method has three phases, *Design DICM-model*, *Refine DICM-model in use*, and *Learn and communicate*. Each of these phases has three steps each. The six steps of the QIP are articulated into nine steps in the DRD-method. The QIP is a six step quality improvement paradigm which is based on GQM and the PDCA.

*Activity 4: Demonstration*
The proposed DRD-method has not been demonstrated to justify that it has a design that exemplifies the solution to the identified problem. However, the major components of the DRD-method, the three phases, and its steps are expressed as valid by all experts who evaluated the method. Therefore, the demonstration is left for future research.

*Activity 5: Evaluation*
A preliminary ex-ante evaluation was conducted by using semi-structured interview of six experts. These experts evaluated the validity of the components of the proposed DRD-method and indicated future directions. The interview was analyzed thematically to identify areas of improvement. These evaluations are used to demonstrate the applicability of the proposed method. The components of the proposed DRD-method were validated using ex-ante evaluation by communicating the method and interviewing six experts. The DRD-method has two cycles – organizational learning and project learning – where the first one deals with the use of previous experiences to design or adapt a DICM-model, and the second one deals with improving DICM-models in use in a similar way as the QIP. Additionally, the GQM and the BSc are included in the proposed DRD-method to design relevant measurement models.

*Activity 6: Communication*
The DRD-method was presented to six experts for evaluation, and it was also published in an international workshop.

## Paper III - Framework of barriers constraining developers

This study is a continuation of an investigation by Hjalmarsson et al. 2014. The investigation by Hjalmarsson et al. 2014 was focused on identifying barriers inhibiting external developers to develop viable digital services in two phases, exploratory and confirmatory phases. During the exploratory phase 15 barriers inhibiting open data service development were identified. In the confirmatory phase, an interview was designed to collect data about barriers and to analyze barriers which lead to the identification of three more new barriers.

The authors used Tavelhack 2013 as a single case study. Travelhack 2013 is chosen as a case for identifying barriers inhibiting the development of viable services from prototypes after contests. Also, because of its temporal nature which enabled the identification of barriers inhibiting the development of market-ready digital artifacts. The same case study is applied in this longitudinal study.

*Activity 1: Exploratory phase*
The exploratory phase is done by Hjalmarsson et al. 2014 as discussed above. The assumption was that barriers constraining the post-process after innovation contests might vary over time in importance, affect each other, and if one barrier is managed then, other barriers will appear.

*Activity 2: Confirmatory phase*

**Choice of case study**
The case, Travelhack 2013 is applied for this study for the same reason as in the Hjalmarsson et al. 2014 study, and for conducting a longitudinal study about post-contest open data innovation barriers.

**Design of survey**
A survey about 18 barriers identified by Hjalmarsson et al. 2014 was designed to collect data with closed questions to rate how barriers are perceived by external developer teams and open-ended questions to identify new barriers, which resulted in six innovation barriers. Besides, comparison of perceived innovation barriers was conducted based on team progress.

**Analysis**
The arithmetic mean and the standard deviation were used as descriptive statistics (Balnaves and Caputi 2001), together with the effect size to examine barrier perception variation over time and analyze differences in perceived barriers 2 and 18 months after the contest. Effect size is an easy way to numerically quantifying the significance of change between groups. It has many benefits over the use of tests of statistical significance, such as hypothesis testing. Effect size effect size, which is not a sample size (Coe 2002), is sometimes preferred instead of hypothesis testing like the t-test because hypothesis testing has problems such as:

1. Statistical power depends on how big the effect is and hence depends on the sample size. For example, larger sample size investigations are characterized by less sampling error and better approximation of the population.
2. The measure of significance is also reliant on sample size. A smaller sample size may lead to loss of important effects, and a larger sample may lead to eliciting very small and insignificant effects to turn out to be significant (Field 2013). In Paper III there are only 17 and 19 responses, which is inadequate for running testing and hypothesizing the characteristics of the groups.
3. The strictness in deciding that an effect is significant, like in hypothesis testing, depends on sample size.

The null hypothesis is forever false since it is based on probability (Cohen 1994, Field and Wright 2006). Statistical significance test, the null hypothesis test might yield insignificant. However, null hypothesis testing does not tell us about the significance of the effect, but Cohen's d does (Field 2013). Therefore, effect size was used to measure and compare change in perceived barriers by developers. Besides, statistical significance depends on sample size and effect size while effect size does not depend on sample size (Sullivan and Feinn 2012) we used effect size to evaluate perceived barriers. The interpretation of Cohen's d, effect size, value is d=0.2 small, d=0.5 medium, and d=0.8 large effect (Field 2013).

# Paper IV - Framework of adopting cloud-based digital services

*Choosing appropriate method*
Due to limited data collection time and accessibility of 290 potentials respondents, quantitative method with survey was chosen. However, it is possible to conduct this study using qualitative approaches.

*Design of survey*
Census, investigation of a finite set, encompasses the whole population under observation. Additionally, in cases where a subset of a population is observed, the study is called sample survey (Särndal et al., 1992). Complete set of the population for this study is available at the Swedish Association of Local Authorities and Regions website (Sveriges Kommuner och Landsting). The respondents of the survey were professionals using cloud-based information systems. The respondents of the survey were 290 Swedish municipalities, and 115 municipalities replied yielding a response rate of 40 %. To validate the adequacy of the number of respondents' sample size formula by Bartlett et al. is used:

According to Bartlett, if continuous data is assumed and if categorical variable plays a primary role in data analysis, then a categorical sample size formula should be used. On the contrary, if a categorical variable doesn't play a primary role in data analysis the sample size formula for continuous data is used (Bartlett et al. 2001).

In this investigation, the study chosen is descriptive statistics and is not categorical analysis to summarize data and the variables. The variables used are implications which are rated with a five point Likert scale for mean and standard deviation calculation. Therefore, the variables in this investigation are not used for categorical analysis and hence sample size determination for continuous data calculation was used as suggested by (Bartlett et al., 2001).

The sample size following Bartlett et al. 2001 is calculated as 83.

In this study, 115 respondents responded and since the minimum sample size calculated is less than 115 it is adequate for this investigation besides larger sample sizes are more probably lead to mean values that resembles the population mean values (Well et al. 1990).

*Collection of data*
A web-based survey using Google survey was designed, and link to the survey was sent via email along with invitation letter to collect evaluation data. The survey includes all implications, opportunities and challenges, listed in a way that respondents will be able to rate the significance of each implication. Additionally, open-ended questions were included to give opportunities for suggesting new implications. Web-based form reduces costs of data collection, saves time, and avoids transcription errors (Andrews et al. 2003).

*Descriptive statistics*
The descriptive statistics was generated using SPSS statistical package. Descriptive statistics provides a summary of the evaluation in terms of mean, variance and standard deviations. In addition, exploratory factor analysis was also applied to identify hidden constructs. Exploratory factor analysis was chosen because the nature of the variables, the relationship among implications and possible hidden constructs, is not known at the beginning.

Williams et al. describe five steps to undertake exploratory factor analysis in SPSS (Williams et al.2010).These steps are automated using SPSS and the result is illustrated in Figure 4.

# Trustworthiness of the result

In this section, the four aspects of trustworthiness such as credibility, transferability, dependability, and confirmability as presented in (Guba 1981) are discussed in relevance with the four papers presented in this thesis. Credibility is about internal validity, and if there are errors, these errors may lead to interpretation errors. Transferability is about external validity and generalizability about applicability. Dependability is related to consistency, reliability, and stability of the data. Finally, confirmability is about neutrality, objectivity, and more of triangulation (Guba 1981).

Data analysis was done separately by each author to avoid disclosing only positive results in Paper II. Both, Paper I and Paper II are externally validated by evaluating the artifacts presented in the papers by experts.

In papers, Paper III and Paper IV Likert scale is used for collecting data. The Likert scale can be data with ordinal ranked values, and Likert type can be Likert items (Boone and Boone 2012). There is controversy among the scientific community about Likert scale as to how the analysis has to be done. Some scholars argue that Likert scale cannot be used in descriptive statistics because of the inadequacy of sample size, laws of normal distribution, and ordinal scale data cannot be analyzed statistically. On the other hand, there are so many research works since the 1930s that consistently portray that it is valid to conduct research using a Likert scale and descriptive statistics (Norman 2010). Likewise, Sullivan and Artino describe the controversy in their article, but they conclude that it is possible to run statistical analysis and the interpretation would yield a valuable result if the data follow a normal distribution (Sullivan and Artino 2013).

However, Boone and Boone argue that Likert scale data can be analyzed with descriptive statistics for analysis using mean for central tendency measure, the standard deviation for variability, person's r for associations, and other statistical methods such as ANOVA, t-test, and regression. On the other hand, Likert type data can be analyzed with median or mode for central tendency measure, frequency for variabilities, Kendall tau B or C for associations, and chi-square for other statistics (Boone and Boone 2012).

In paper IV, the sample size was calculated using a sample size formula specifically designed with an assumption of continuous data. Thus the sample size was calculated using the formulas proposed by (Bartlett et al. 2001). Additionally, the adequacy of the sample size and the appropriateness of the data for factor analysis is made using SPSS, and it has reasonable KMO (Kaiser-Meyer-Olkin) which indicates that the sample size is enough for factor analysis according to (Williams et al.2010).

## Ethical considerations

Confidentiality of respondents' data is well addressed according to the privacy and confidentiality issues discussed by Andrews et al. (2003). Ethical issues for qualitative research according to Creswell is also is addressed in the process of investigation: before conducting the study, beginning to do the study, collecting data, analyzing data, reporting data and publishing the work (Creswell 2013). For example, research authorship negotiation has been done and during the study relevant issues are addressed such as informing respondents the purpose of the survey. Also, during collection of data the purpose of the study and how data is collected is explicated to participates, and privacy of respondents' is addressed by aliasing their identity in publications. Finally, reporting of the results is published to share the result.

# 4. Results

In this chapter, the results of the papers, that are the basis for this thesis, are presented. The results are divided into two: i) research findings about the evaluation of the development of digital services from open data stimulated by contests and post contest development and ii) research findings of evaluating barriers constraining external developers to enter the market and the adoption of digital services over the cloud to improve their utilization.

## Evaluation of Contest-Driven Digital Service Development and the Succeeding Deployment to Viable Service

### Paper I – DICM-model

**Model for measuring digital innovation contests (DICM-model)**

In Paper I, a list of innovation evaluation models and frameworks are reviewed. Few of the existing models or frameworks reviewed measure innovation contests, and none of the models or frameworks are designed to evaluate digital innovation contest to stimulate open data development. For example, some of these models and frameworks are designed to evaluate organizational innovation, product innovation, process innovation, and innovation in nations. Furthermore, the evaluation of the post-contest process is overlooked in existing literature. Therefore the purpose of Paper I is to design a model, the DICM-model, for evaluating open data service development stimulated by contests and the post contest process. Therefore, the measurement model for digital innovation contests, DICM-model, is proposed to address this gap and to introduce an evaluation model for digital innovation contests and service deployment processes.

The design of the model is based on design science approach, and therefore the six steps of Peffers et al. (2007) are followed.

The problem identification is related to the evaluation of innovation. Evaluating the effects of innovation contests is a challenging endeavor. Accordingly, to gain an improved understanding of the problem an exploratory case study was used. A literature study was done to elicit the objectives of the solution.

The objectives of the solution are to:
- support in measuring goal fulfillment of an innovation contest

- enable organizers in identifying strengths and weaknesses in the IVC
- aid organizers in learning and increasing maturity
- be easy to apply by organizers of innovation contests.

The design and development of the proposed model, such as identification of the main characteristics and activities are based on a case study.. Also, the measurement model is applied to the case study for the demonstration. The come of the evaluation of the DICM-model indicated that there is a need for a more agile method for designing, adapting, and refining the DICM-model to be applicable for a given contest.

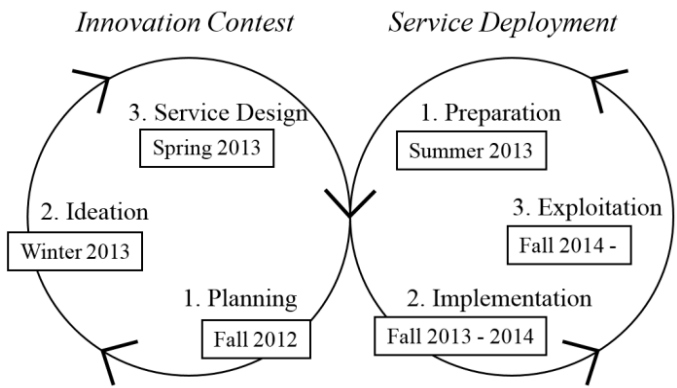

*Figure 2:The innovation contest process and the service deployment processes.*

The DICM-model measures two processes of digital innovation contest: innovation contest and service deployment. The DICM-model adapts the three phases of the innovation value chain of Hansen and Birkinshaw (2007) to include perspectives of contests in each process as illustrated in Figure 2. For example, innovation contest includes planning, ideation, and service design; and service deployment includes preparation, implementation, and exploitation. The components of the phases of the DICM-model are based on the evaluation model by Erkens et al. (2013) both models inputs, activities, outputs, and measures refer to Table 4 and Table 5 below.

The three phases of the innovation contest process have specific objectives. Planning is the phase where organizers design contests, ideation is the phase where organizers support idea generation and select finalist, and finally, service design is a phase by which prototypes are developed from ideas. Each of these phases has inputs, activities, outputs, and measures

| Phase | Planning | Ideation | Service Design |
|---|---|---|---|
| Input | Resources, for example API info, open | Time, resources and facilities | Time, resources and facilities |
| | data sources, domain knowledge, financial resources | | |
| **Activities** | a. Specify problem – solution space<br>b. Design contest, i.e. applying the design elements, establish evaluation criteria<br>c. Market contest, i.e. events, website, media coverage, make resources available | a. Support in idea generation, e.g. problem descriptions, personas, meet-ups, technical support, business model support<br>b. Select finalists: evaluate ideas and business models | a. Support in service design, e.g. hackathon, technical support, business model support<br>b. Select winners: evaluate prototypes and business models |
| **Output** | Registered participants ready to contribute to the competition | High quality digital service ideas | High quality digital service prototypes |
| **Measures** | • Available resources<br>• Problem – solution maturity<br>• Contest quality<br>• Visibility<br>• Number of participants | • Available resources<br>• Utilization of available resources<br>• Problem - solution maturity<br>• Quality of support<br>• Time invested by participants<br>• Number of submitted ideas<br>• Ratio of ideas per participant<br>• Number of high quality digital service ideas<br>• Visibility | • Available resources<br>• Utilization of available resources<br>• Problem - solution maturity<br>• Quality of support<br>• Time invested by participants<br>• Number of digital service prototypes<br>• Ratio of prototypes per participant<br>• Number of high quality digital service prototypes<br>• Visibility |

*Table 4: Measurement model for the Innovation Contest Process.*

The service deployment has three phases by which organizers based on a chosen level of involvement are involved in service deployment through preparation, implementation to convert the prototypes to viable digital service, and finally exploitation of the market potential.

| Phase | Preparation | Implementation | Exploitation |
|---|---|---|---|
| **Input** | Resources, such as open data, knowledge, relationships, time and money. | Time and resources depending on level of post-contest support | Time and resources depending on level of post-contest support |
| **Activities** | a. Decide level of post-contest support<br>b. Establish goals for service deployment<br>c. Organize resources based on goals (in a)<br>d. Go/No go decision | a. Support service implementation at various levels (from no support to very high support)<br>b. Evaluate service quality<br>c. Evaluate market potential<br>d. Go/No go decision | a. Support service delivery at various levels (from no support to very high support)<br>b. Support service commercialization at various levels (from no support to very high support)<br>c. Continuous evaluation of service quality and market potential |
| **Output** | Prepared organization | Viable digital service, business model and intellectual property | Service revenue |
| **Measures** | • Level of post-contest support<br>• Available resources<br>• Level of commitment | • Available resources<br>• Quality of support<br>• Problem – solution maturity<br>• Service demand | • Available resources<br>• Quality of support<br>• Problem – solution maturity<br>• Service usage<br>• Rate of diffusion<br>• Number of downloads |

| | | | • Revenues |

Table 5: Measurement model for the Service Deployment Process.

## Paper II – DRD-method

**Method for designing DICM-models**

In Paper II, a method for designing and refining DICM-models for organizers, the DRD-method, is proposed. The need for the DRD-method arose after evaluating the DICM-model by 13 experts, the DICM-model was designed based on a single case study and hence it doesn't evaluate the fulfillment of goals of contests and service deployment processes for a given contest. Therefore, the proposed method, DRD-method, is designed following the QIP in which the Goal-Question-Metric (GQM) is used to illicit measures. Additionally, the elicitation of measures is enhanced to include strategic perspectives by including the Balanced Scorecard (BSc) to elicit measures. Hence, the components of the DRD- method enable agility in measuring the fulfilment of goals of innovation contests. Also, the method facilitates knowledge management to refine, record and communicate best practices. Six experts also evaluate the DRD-method.

The design of the DRD-method is illustrated in Figure 3. The components are based on QIP of Basili et al. (1994), and by evaluation. The QIP's six steps are adopted to design the DRD-method. For example, the first three steps in the QIP of NASA by Basili et al. (1994), such as *Characterize*, *Set goals*, and *Choose process* are articulated and mapped as *Characterize*, *Set measurement goals*, and *Build measurement model* respectively in Phase 1 using GQM and BSc as measure eliciting techniques. The fourth step in QIP, *Execute*, is adopted into three steps in the new method such as *Analyze result*, *Measure*, and *Provide immediate feedback* in Phase 2. The fifth step in QIP, *Analyze* is mapped to *Analyze* in the third phase. Finally, the sixth step, Package, in QIP, is articulated and mapped into two steps such as *Package* and *Disseminate*.

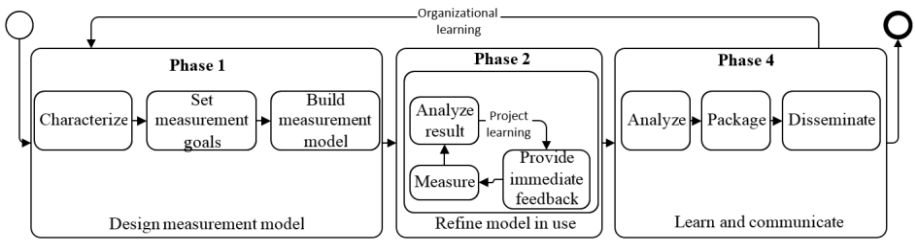

*Figure 3: A nine-step method to design and refine DICM-models*

**Phase 1: Design measurement model**

In this phase, organizers carry out activities listed in steps 1, 2, and 3 to design a DICM-model for a specific contest
.
**Step 1. Characterize**

In this step, organizers carry out elicitation of contest requirements, understand contest goals, and design processes. Knowledge bases can be used to reuse or adapt previously used models. Theoretical foundations are also utilized to characterize innovation contests and post contest deployments.

**Step 2. Set measurement goals**

Organizers identify relevant perspectives and list goals specified by contest owners, specify contest goals if goals are not specified, and articulate goals into sub-goals to facilitate the evaluation of fulfillment. Organizers can also categorize these goals under perspectives according to BSc. Finally, questions are articulated to measure evaluation of contest goal fulfillment.

**Step 3. Build measurement model**

Organizers identify, define and describe processes, phases, inputs, activities, outputs and measures of their DICM-models based on characteristics and questions. Since questions are derived from goals and metrics from questions, the GQM paradigm is an important mechanism for building the measurement model. Identification of required data sources is also an important activity when building a measurement model. Qualitative measures such as market potential can be assigned scales from 1 to 5 with quantitative ordinal values with labels: very insignificant, insignificant, neutral, significant, and, very significant correspondingly.

**Phase 2: Refine model in use**

In this phase, organizers assess contests and service deployments by measuring current processes, analyzing the result and providing immediate feedback. The feedback can be used to improve performance and make modifications to the model. This Phase is an iterative phase which terminates when organizers decide to terminate the contest.

**Step 1. Measure**

In this step, organizers set timeline and start contests for each phase of the DICM-model. Organizers follow the timeline to execute activities in each phase of the DICM-model using inputs to produce outputs and collect measures to provide execution feedback.

**Step 2. Analyze result**

In the second step, organizers analyze collected measures to identify deviations and their causes. Also, organizers suggest coping strategies for barriers encountered.

**Step 3. Provide immediate feedback**

In this step, organizers combine and communicate best practices and measured performances to suggest improvements and strategies to cope up with challenges. Also, the current DICM-model is refined by organizers to reflect current contest situation. For example, if feedbacks are indicating that there are new inputs, activities, and or outputs, then the needs are incorporated in the current measurement model.

**Phase 3: Learn and communicate**

This phase guides the process of management of knowledge by storing best practices after identifying problems collected during measurement to identify lessons learned. These best practices and experiences gained are stored in the knowledge base and communicated for future purposes and scientific community.

**Step 1. Analyze**

Organizers, after completion of the contests, analyze data regarding measurements, analyze used practices to identify problems, archive findings and make a recommendation of best practices.

**Step 2. Package**

Organizers of contests update their knowledge base. Knowledge bases are updated with information regarding best practices from using customized DICM-model or new DICM-model. Organizers combine best practices gained from using any of these models for future projects.
.
**Step 3. Disseminate**

Lessons learned from contests are disseminated and communicated for practice and science. .

# Evaluating barriers constraining external developers to enter the market and the adoption of cloud-based services to improve the utilization

Paper III – Framework of Innovation Barriers

**Framework of innovation barriers constraining open data service development**

In the research area of open data service field, how sustainable open data service development is carried out is an area of investigation. In Paper III, innovation barriers to open data service development by Hjalmarsson et al. are evaluated as perceived by developers through the course of development. Hjalmarsson et al. identified 18 barriers from a literature review and empirical investigation of innovation contest participants (Hjalmarsson et al., 2014). To better understand how barriers vary in importance over time and the dynamic nature of innovation barriers Paper III presents an investigation of existing barriers and identified new barriers. Therefore, these barriers as perceived by external developers constraining those in need of developing viable services were measured with longitudinal survey by comparing 2 months vs. 18 months post-contest evaluations as shown in Table 6. Also, during the survey open-ended questions were additionally provided to respondents to identify new barriers and six new barriers are identified, B19, B20, B23, B22, B23, and B24, see table below.

| Barriers to Open Data Service Development | Post-contest evaluation results | | | | Comparison |
|---|---|---|---|---|---|
| | 2-months | 18-months | | | |
| | Mean | Mean | Median | St. Dev. | Tendency Changes in mean |
| B3. Lack of time or money to prepare market entry | 4,32 | 4,00 | 5 | 1,56 | ↘ |
| B22. Lack of model to generate revenues to sustain the service | N/A | 4,00 | 4 | 1,09 | ↑ |
| B23. Lack of interest within the team to pursue development | N/A | 3,82 | 5 | 1,44 | ↑ |
| B11. Weak value offering | 3,21 | 3,71 | 4 | 1,39 | ↘ |
| B20. Needed open data sets are missing | N/A | 3,53 | 4 | 1,66 | ↑ |
| B5. Lack of external funding | 2,11 | 3,53 | 4 | 1,62 | ↗ |
| B6. Multifaceted market conditions and uncertain product demand | 2,84 | 3,35 | 3 | 1,22 | ↗ |
| B19. Lack of quality in used open data | N/A | 3,29 | 4 | 1,44 | ↑ |
| B14. Difficulties in reaching adequate technical quality in the service | 1,89 | 3,18 | 4 | 1,59 | ↗ |
| B21. Changes in used APIs at short notice | N/A | 3,12 | 3 | 1,30 | ↑ |
| B24. Hard to interact with data providers | N/A | 3,12 | 4 | 1,44 | ↑ |
| B18. Hindering industry structures | N/A | 3,12 | 3 | 1,30 | ↑ |
| B10. Lack of partner co-operation for technical development | 3 | 3,06 | 4 | 1,47 | → |
| B7. Lack of marketing competence and market information | 3,26 | 3,00 | 3 | 1,39 | ↘ |
| B4. High market competition and saturation | 1,84 | 2,88 | 3 | 1,54 | ↗ |
| B17. Viable product features uncertainty | N/A | 2,53 | 3 | 1,45 | ↑ |
| B15. Lack of partner co-operation for technical tests | 1,79 | 2,24 | 1 | 1,45 | ↗ |
| B13. Varieties of smartphones requiring unique service development | 2,42 | 2,12 | 1 | 1,41 | ↘ |
| B12. Limitations in existing service-dependent platforms | 1,53 | 1,94 | 1 | 1,18 | ↗ |
| B9. Difficulties establishing licenses for APIs and other services | 1,95 | 1,76 | 1 | 1,24 | ↘ |
| B16. Lack of partner co-operation for knowledge transfer | N/A | 1,76 | 1 | 1,29 | ↑ |
| B1. Lack of technical competence and innovation experience | 1,84 | 1,41 | 1 | 0,79 | ↘ |
| B2. Difficulties finding competent team members | 1,32 | 1,29 | 1 | 0,68 | → |
| B8. Inefficient intellectual property processes | 2 | 1,18 | 1 | 0,73 | ↘ |

*Table 6: Perceived barriers in order of importance from highest (score 5) to the lowest (score 1)*

A comparison of perceived barriers affecting team progress was also investigated to learn if there are barriers perceived vary among teams. This comparison is presented in the table below, Table 7.

| Barriers to Open Data Service Development | Comparison of mean based on team progress after the contest | | |
|---|---|---|---|
| | 10 to 18 months | 2 to 9 months | 1 month or less |
| B3. Lack of time or money to prepare market entry | 1,67 | 4,57 | 4,43 |
| B22. Lack of model to generate revenues to sustain the service | 4,33 | 4,14 | 3,71 |
| B23. Lack of interest within the team to pursue development | 1,67 | 4,00 | 4,57 |
| B11. Weak value offering | 3,00 | 3,71 | 4,00 |
| B5. Lack of external funding | 3,67 | 3,86 | 3,14 |
| B20. Needed open data sets missing | 5,00 | 2,71 | 3,71 |
| B6. Multifaceted market conditions and uncertain product demand | 3,33 | 3,14 | 3,57 |
| B19. Lack of quality in used open data | 4,33 | 3,43 | 2,71 |
| B14. Difficulties in reaching adequate technical quality in the service | 4,67 | 3,29 | 2,43 |
| B18. Hindering industry structures | 4,33 | 3,14 | 2,57 |
| B21. Changes in used APIs at short notice | 4,33 | 2,71 | 3,00 |
| B24. Hard to interact with data providers | 3,00 | 3,29 | 3,00 |
| B10. Lack of partner co-operation for technical development | 3,33 | 3,00 | 3,00 |
| B7. Lack of marketing competence and market information | 2,00 | 3,57 | 2,86 |
| B4. High market competition and saturation | 3,33 | 2,71 | 2,86 |
| B17. Viable product features uncertainty | 2,33 | 2,29 | 2,86 |

| B15. Lack of partner co-operation for technical tests | 3,33 | 2,57 | 1,43 |
| B13. Varieties of smartphones requiring unique service development | 1,00 | 2,14 | 2,57 |
| B12. Limitations in existing service-dependent platforms | 2,00 | 2,00 | 1,86 |
| B9. Difficulties establishing licenses for APIs and other services | 1,00 | 1,71 | 2,14 |
| B16. Lack of partner co-operation for knowledge transfer | 3,00 | 2,00 | 1,00 |
| B1. Lack of technical competence and innovation experience | 1,33 | 1,71 | 1,14 |
| B2. Difficulties in finding competent team members | 1,00 | 1,29 | 1,43 |
| B8. Inefficient intellectual property processes | 1,00 | 1,43 | 1,00 |

*Table 7. Comparison of perceived barriers in order of importance based on team progress*

This paper adds a third phase to the research process reported in Hjalmarsson et al. (2014) transforming it into a longitudinal survey. The two phases are activation and building development momentum. In this paper, a comparison about perceived barriers after 2-months and 18 months of survey is done. In this third phase, the *comparison* phase, data was collected in two iterations.

The results in this paper show that external developers face post-contest innovation barriers in at least three phases. These phases are *activation*, *building development momentum* and *preparing for market entry*. In the first phase, *activation*, external developers mainly strive to achieve to mobilize for free time or financial resources to be able to prioritize continued development. At this stage developers also try to understand the market demands on future open data services that intended customers have, and make the service attractive for stakeholders.

Developers that can activate and pass activation enter the *building development* phase where they build development momentum, the second phase. The result of the survey indicates that external developers who focus on contributing to the development two months after the contest have a higher chance of entering the second phase. External developers face two main barriers, time and resources, in both the first and the second phases. Besides, in the second phase, developers also face barriers such as lack of external funding and lack of a revenue model to create resources for maintaining the service after finalization. If developers can progress the development in this phase, then they may face lack of quality in open data and challenge to guarantee quality in their service.

The third phase is *preparing for market entry*. If the *development momentum*, the second phase, is built, then developers enter the third phase. This phase is mainly concerned with having more focus on guaranteeing the quality of the features selected for the service to be launched. At this phase, there are few barriers constraining development.

# Adoption of Cloud Based Digital Services

## Paper IV – Framework of Service Adoption

**Guidelines for adopting cloud based digital services**

The objective of Paper IV is to empirically evaluate implications of cloud-based digital services in the public sector. Juell-Skielse and Enquist provide a framework of more than 80 implications, opportunities & challenges, from the perspectives of users and suppliers (Juell-Skielse and Enquist 2012). However, these implications were not evaluated by users and providers of cloud computing, and the framework cannot be used to aid in the adoption of cloud computing with significance in mind. Therefore, in Paper IV user implications are evaluated from the perspectives of cloud-based digital service users in the public sector. These implications are prioritized using descriptive statistics as illustrated in Table 8 and Table 9 and Figure 4.

Software-as-a-Service (SaaS) is one of the deployment models of cloud computing (NIS). Cloud computing provides new opportunities for innovation, marketing opportunities for companies and efficiency gains (Kushida et al. 2011). Furthermore, public information systems are needed to fulfill various purposes such as health-care, financial sectors, school administration and geographical data to the public delivered by the authorities and municipalities (White 2007). Even if on premise installations of standard packages or custom-built software are common in municipalities, cloud-based public information systems delivered as service provide several advantages for public organizations (Diez and Silva 2013). Despite the interest in cloud computing studies in implications, opportunities & challenges, provide an unstructured list of implications which is not directly applicable for adopting cloud-based systems as stated in Paper IV. Therefore, there is a need for analysis of the significance of implication.

The evaluation of implications is carried out through a survey of Swedish municipalities, out of 290 municipalities 115 replied. Descriptive statistics is used to quantitatively summarizing the collected data into lists of prioritized implications as illustrated in Table 8 and Table 9. Also, a statistical analysis technique called exploratory factor analysis the number of implications is reduced by grouping them into factors. These factors are also known as hidden constructs that can explain all the variables, implications, by grouping them as representational implications see Figure 4.

| Rank | Opportunities | Mean | Std. Dev. |
| --- | --- | --- | --- |
| 1 | Remote access from anywhere at anytime | 3.63 | 1.013 |
| 2 | Access to and flexibility to choose between state of the art technologies | 3.57 | 0.909 |
| 3 | Easier version management | 3.53 | 0.862 |
| 4 | Predictable costs | 3.35 | 0.956 |
| 5 | Access to reliable, secure and scalable infrastructure | 3.20 | 0.929 |
| 6 | Decreased implementation costs | 3.19 | 0.926 |
| 7 | Reduced up-front investments | 3.14 | 0.897 |
| 8 | Increased standardization | 3.13 | 1.064 |
| 9 | More focus on IT-value | 3.08 | 1.036 |
| 10 | Customer responsiveness | 3.07 | 1.032 |
| 11 | Simplified phasing of implementation | 2.92 | 0.938 |
| 12 | Improved processes | 2.91 | 0.913 |
| 13 | Productivity improvements | 2.90 | 0.917 |
| 14 | Increased focus on core competencies | 2.89 | 1.049 |
| 15 | Order cycle improvements | 2.85 | 0.891 |
| 16 | More proactive purchasing behavior | 2.83 | 0.985 |
| 17 | Lower costs | 2.82 | 0.884 |
| 18 | Improved information and transparency | 2.76 | 0.904 |
| 19 | Single Point of Contact | 2.75 | 1.146 |
| 20 | Personnel reductions | 2.75 | 0.916 |
| 21 | Easier access to technical expertise | 2.74 | 1.069 |
| 22 | Decreased implementation risks | 2.72 | 0.969 |
| 23 | Increased bargaining power | 2.69 | 0.912 |
| 24 | Improved financial close cycle | 2.61 | 0.876 |
| 25 | Increased integration of information | 2.57 | 1.101 |
| 26 | Complete service offerings from several vendors | 2.37 | 1.054 |

Table 8: Ranked list of user opportunities with cloud based digital services for the public

| Rank | Challenges | Mean | Std. Dev. |
| --- | --- | --- | --- |
| 1 | Large dependency on vendor | 3.83 | 0.993 |
| 2 | Less customization possibilities | 3.49 | 0.968 |
| 3 | High demands on process orientation | 3.39 | 0.915 |
| 4 | Less integration possibilities | 3.39 | 1.09 |
| 5 | Increased security risks | 3.37 | 0.959 |
| 6 | Redistribution of responsibility | 3.34 | 0.887 |

| 7 | High demands on IT-projects to be business initiatives | 3.33 | 1.057 |
| 8 | Lack of policies and laws | 3.3 | 1.019 |
| 9 | Project escalation and lack of control | 3.14 | 1.083 |
| 10 | Lack of project team expertise | 3.13 | 0.903 |
| 11 | Lack of senior management involvement | 3.1 | 1.116 |
| 12 | Structural changes | 3.07 | 0.915 |
| 13 | Poor use of consultants | 3.04 | 1.021 |
| 14 | Lack of detailed systems implementation plan | 3.03 | 0.945 |
| 15 | Software-as-a-service requires local software | 2.99 | 1.158 |
| 16 | Lack of involvement of internal audit | 2.82 | 0.923 |
| 17 | Less availability, reliability and performance | 2.73 | 0.809 |
| 18 | More rigid organizations | 2.63 | 0.986 |
| 19 | User resistance | 2.59 | 0.981 |

*Table 9: Ranked list of user challenges with cloud digital services for the public sector.*

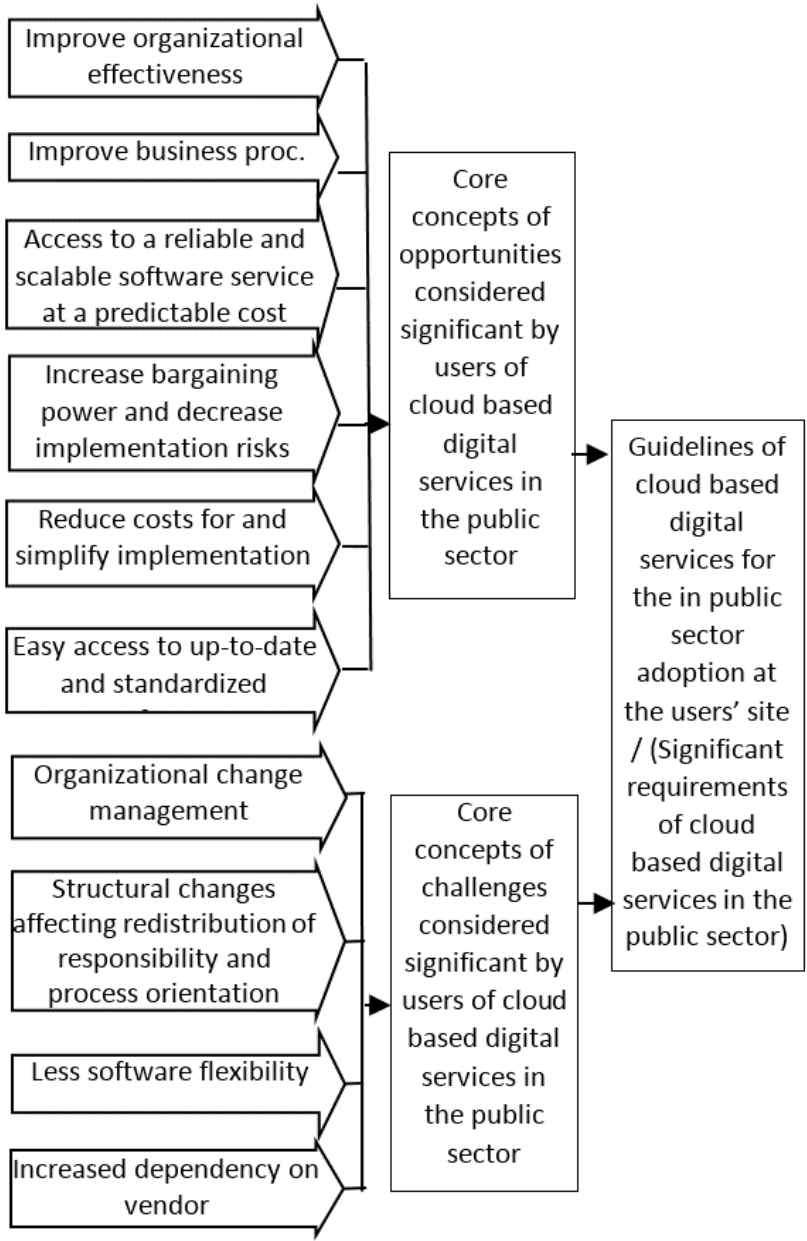

*Figure 4: Modified framework of cloud based digital services adoption.*

The modified framework, the prioritized list of implications and the result of the exploratory factor analysis, can be used to adopt cloud-based digital services which in turn facilitate the utilization of digital services.

# 5. Discussion

Innovation in open data digital service development stimulated by contests, the succeeding service deployment and the adoption of cloud-based digital services improves the utilization of digital services. The four papers, Paper I to Paper IV discuss ways of improving the utilization of digital service through contests, digital service development, and adoption of cloud-based digital services. In this chapter, the two research questions discussed in the Introduction, open data digital service developments stimulated by contests and the preceding service deployment with barriers, and cloud-based digital service adoption are discussed. Finally, limitations are also presented.

## Research Challenges

Open data has various benefits to the society, and it is becoming common means for advancing digital service development by unlocking the prospects such as improving services and entrepreneurship (Lakomaa and Kallberg 2013). Open data service development, however, is still in its infancy (Janssen et al. 2012). Also, some open data services developed by contests do not reach the market due to barriers faced by external developers (Hjalmarsson et al. 2014).

On the other hand, scalability and speed can be achieved by adopting cloud-based digital services (Armbrust et al. 2010. It is also evident that cloud computing enables innovation and entrepreneurship (Kushida et al. 2011). All these benefits may lead to the adoption and the spending in the private and the public sector. For example, IDC estimated that cloud spending would reach 195 billion dollars in the public sector worldwide by 2020 (IDC, 2016). Similarly, the adoption of cloud-based digital services in the private sector transforms the life cycle of product development (Wu et al. 2016) and empowers users and suppliers with opportunities (Juell-Skielse and Enquist 2012). However, existing implications are not prioritized and structured to guide the adoption of cloud-based digital services.

Research challenges presented in the previous two paragraphs such as the availability of open data having a number of benefits while the utilization it still at its infancy. For example, the use of open data service development stimulated by contests enables the utilization of open data to develop digital services; but a limited number of services reach the market due to barriers faced by external developers. Moreover, opportunities of the cloud-based digital services such as scalability, innovation, market prospects of adopting the cloud, and others despite the availability of un-prioritized and unstructured framework of implication for adoption. These challenges are reflected in the research question: "*How can contest-driven innovation of open data digital services be evaluated and the challenges constraining the development and the adoption of cloud-based digital services be supported to improve the utilization of digital services?*" This research question is further divided into two questions RQ1: "*How can the process of contest-driven digital innovation be evaluated?*" and RQ2: "*What are the barriers constraining the development of digital services, and the challenges to effectively adopt cloud-based digital services?*"

# Development of Contest-Driven Digital Services Using Open Data

The first research question, RQ1, is addressed by two papers, Paper I and Paper II. Artifacts proposed in Paper I and Paper II the DICM-model and the DRD-method respectively deal with the evaluation of the development of digital services using open data through evaluation of innovation contest. The DICM-model is used to evaluate digital innovation contest processes while the DRD-method enables to design and refine digital innovation contests and the following service deployment processes. These two papers are discussed below.

The DICM-model, Paper I, adds a theoretical and empirical anchored model, the DICM-model, to measure digital innovation contests and post-contest deployment processes. The model provides support to evaluate digital innovation contests from beginning to end. The model increases the knowledge about how contests affect the Innovation Value Chain from the viewpoint of contest organizer and provide a detailed cross-check for more general models. Available measurement models are based on IVC (Ishak et al. 2014; Ress et al. 2013; Hansen & Birkinshaw 2007). Similarly, the DICM-model is also based on IVC. IVC enables the identification of strengths and weaknesses in the IVC (Hansen & Birkinshaw 2007).

Paper II proposes a method for designing and refining DICM-models, DRD-method. Some measurement models available in literature do not provide flexibility to adapt to contexts that vary from organizer to organizer. For example, the model by Hansen & Birkinshaw is limited in the number of measures in each phase (Hansen & Birkinshaw 2007). However, the model by Misra et al. is flexible since measures are derived from goals (Misra et al. 2005). Similarly, the DICM-model lacks flexibility probably attributed to the fact that it is based on a single case study and literature review. Therefore, a goal-oriented approach was suggested to build or adopt DICM-models.

The model by Misra et al. elicits measure from goals and it is based on GQM paradigm. The GQM paradigm is introduced in the software engineering discipline (Basili and Weiss 1984) but is becoming popular in healthcare (Villar 2011), information security (Kassou and Kjiri 2013), and information systems (Esteves et al. 2003). The DRD-method is inspired by the QIP which is, in turn, builds on GQM and the PDCA. Additionally, BSc is also included in the DRD-method to identify strategic objectives. The combination of the BSc and the GQM is productive since both cover different aspects of measures despite their similarities in the way they derive measures (Buglione and Abran 2000). The QIP assures continuous process improvement (Basili et al. 1994).

# Barriers Constraining Digital Services Development and the Adoption of Cloud-Based Digital Services

Finally, the second research question, RQ2, is addressed by paper III and Paper IV. In Paper III barriers constraining external developers hindering the development viable digital services is presented. Furthermore, in Paper IV the framework of implications of cloud-based digital services in Paper IV is empirically assessed to prioritize them based on significance, opportunities & challenges, using descriptive statistics. Juell-Skielse and Enquist developed a framework of implications of adopting cloud-based digital services provisioned in SaaS service model from users' and suppliers' perspectives (Juell-Skielse and Enquist 2012). Paper IV, presented a prioritized list of implications from users' perspectives and a result of exploratory factor analysis leading to the identification of hidden implications.

The service deployment process is dealing with the implementation and exploitation of viable digital services and barriers constraining external developers is evaluated in Paper III. Paper III presents a longitudinal survey of innovation barriers limiting external developers after contests from developing viable digital services. These barriers were derived from literature and during empirical inquiries in innovation contests. Innovation barriers are evolutionary, complementary and dynamic hence barriers act at different stages of innovation process (Hadjimanolis 2003). Similarly, the result in Paper III shows that there are variations in the perception of barriers among external developers and barriers are perceived differently through time in three phases such as *activation*, *building development momentum*, and *preparing for the market entry*. In the *activation* phase, developers struggle to understand market potential, allocate time and organize resources. If they can succeed to activate developers, they enter the *development phase* where they phase barriers such as time, external funding, resources, and increased interaction with data providers among others. Developers who could sustain the lack of time or lack of money are not managed then developers will disappear, and the service development will discontinue, and semi-finished service will never enter the market. Whereas, if developers could sustain these barriers, they can enter the third phase, thereby *preparing for the market entry*. In this phase, developers face challenges related to quality, changes in APIs as well as technical barriers. The longitudinal survey of barriers built on previous barriers resulted in 24 barriers, and the analysis indicates that external support for developers to coup with barriers must be carefully provided.

Although the modified framework of implications, Paper IV, is applicable for adoption in the public sector, the original list of implications is a generic list which is not bound to an area (Juell-Skielse and Enquist 2012). Hence it is also useful for cloud adoption in other sectors. Goal oriented requirement engineering is believed to be a promising paradigm for adopting cloud computing for goals that are flexible and generic (Zardari and Bahsoon 2011). The framework of implications of cloud-based services is applicable for requirement elicitation for digital services by articulating important opportunities and challenges into sound goals. Therefore, the modified framework of cloud-based digital services can be used in requirement elicitation for the adoption of cloud computing.

## Limitations

Some of the artifacts developed in the four papers have limitations related to the methods of development, respondents, and applicability.

The limitations of the development processes are related to respondents. For example, respondents of the DICM-model and the DRD-method evaluations are mainly from European countries except for one Asian respondent. Furthermore, finding respondents from another part of the world is not an easy task. Therefore, the result has to be properly validated in different contexts.

Digital service evaluation artifacts, from Paper I and Paper II, are designed for contest organizers. Hence the artifacts are not designed for developers. Furthermore, the artifacts are mainly developed for evaluating digital service development utilizing open data.

The framework of barriers, Paper III, can serve as a starting point for further investigations to manage innovation barriers constraining digital service development utilizing open data. Since contests are events that are occasionally organized the framework of barriers is an indispensable research finding.

However, since this work is based on a single case study, the research work should be properly validated and tested in other contexts and settings.

Finally, the framework of adopting cloud-based digital services is developed from perspectives of users. Hence, it can be applied to procure, customize and adopt cloud-based services. However, it is developed based on a survey of Swedish public municipalities, and further research needs to be done in other sectors to increase its applicabilities.

For example, both the DICM-model and the DRD-method are limited regarding evaluation. The applicability and ex-post evaluation of the DICM-model and the DRD-method should be conducted.

# 6. Conclusions and future research

In this thesis, the improvement of the utilization of digital services through the development of digital services from open data stimulated by contests and the adoption of cloud-based digital services is presented. The contribution includes theory for analysis, explanation and prediction, or more specifically:

- A model for evaluating digital innovation contests, DICM-model
- A method for designing and refining DICM-models, DRD-method
- A framework of barriers constraining external developers from building viable digital services
- A framework of adoption of cloud-based digital services, by providing a prioritized list of implications of adopting cloud-based digital services.

The artefacts developed answer the research question, specifically its two sub questions as illustrated below in Table 10.

| RQs | Artefacts | Potential users |
|---|---|---|
| RQ. 1. | • DICM-model (Paper I) | • Organizers |
|  | • DRD-method (Paper II) | • Organizers |
| RQ. 2. | • Framework of innovation barriers (Paper III) | • Organizers, developers may use list of barriers to prepare for coping strategies |
|  | • Framework for adopting cloud-based digital services (Paper IV) | • Users and Developers |

*Table 10: Artifacts developed to answer Research questions and potential users.*

These contributions, artefacts, listed above have theoretical as well as practical significance. The DICM-model contributes by adding a new measurement model in the open data innovation arena which supports evaluation of DICos. Similarly, the DRD-method adds a new perspective of evaluating DICos, in particular, the innovation contest and the service deployment processes. This method provides a means for knowledge management by enabling organizers to design and refine DICMs. Also, the framework of barriers constraining external developers lists 24 innovation barriers in association to technology, regulation, organization, market, knowledge, and finance. Also, the evaluation of perceived barriers is positioned in association to three post-contest phases namely activation, building development momentum and preparing market entry. Finally, the framework of implications of cloud-based digital services provides a theoretical basis for explaining the comparison of the significance of implications. In addition, the modified framework of implications provides a list of hidden implications that were elicited from existing implication. These implications pro-vide a reduced list of opportunities and challenges which can be used as a guideline of adoption of cloud-based digital services. The artifacts DICM-model and DRD-method are designed following design science research approach. Therefore, the knowledge contribution falls under Improvement for DICM-model and Exaptation for DRD-method in knowledge contribution framework of DSR by Gregor and Hevner (2013) as illustrated in Figure 5 below.

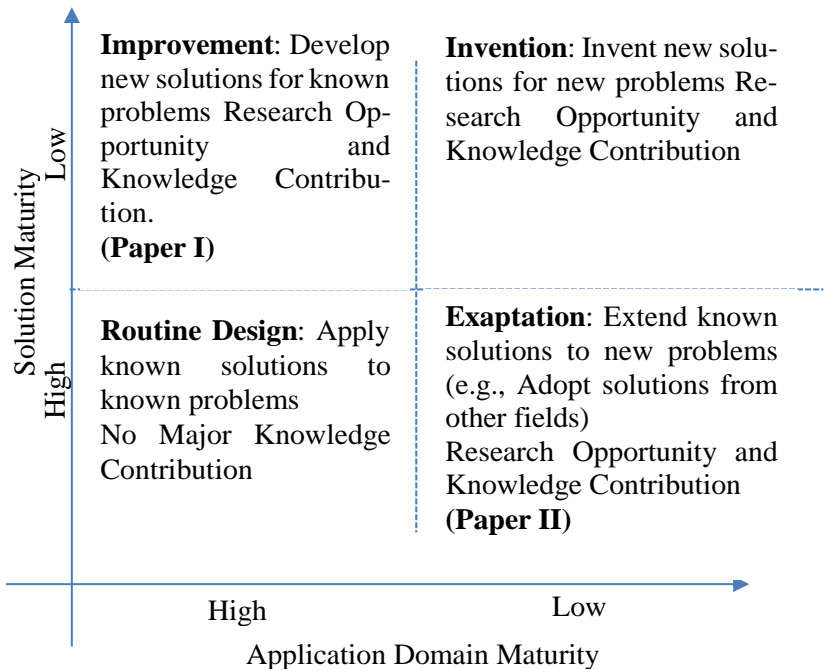

*Figure 5: Knowledge contribution framework of DSR (Gregor and Hevner 2013, page 345).*

To practice, the DICM-model, and the DRD-method adds a new knowledge on how innovation contests and service deployments are managed to enable the development of viable digital services. In particular, these two artifacts enable the evaluation of DICos by setting up goals and measure the fulfillment of goals. Moreover, the DICM-model, due to its recursive nature, enables learning and increasing maturity. Also, the framework of barriers indicated that external support is needed to enable external developers to cope with barriers. Finally, the framework of adoption of cloud-based digital services can be used as guideline for adopting cloud-based digital services.

# Future Research

The integration of the evaluation of digital innovation, DICM-model & DRD-method, in developer portals such as Android, Apple, Microsoft, and others to facilitate the evaluation of digital innovation artifacts is also a potential future research. Most developer platforms do not support the process of innovation in particular.

The first phase of the digital innovation contests is ideation and sometimes planning is done separately before ideation. The rationale behind contests is idea generation to improve the utilization of open data digital service development. Since ideation is the most decisive part of digital innovation, it is significant to undergo a more efficient way of idea generation and trend analysis. Trend analysis is needed to support the identification of relevant ideas for contests and organizations that are inspired by innovating their processes, products and so on. Trend analysis can be done by running network analysis on patent citation (You et al. 2017). The use of web search engine for generating temporal semantic contexts is valuable for applications such as trend analysis (Xu et al. 2014). Also, Piller et al. proposed a typology of co-creation and discussed how social media enables innovation in new product development (Piller et al. 2011). Commercial companies need to devise a way to harness a network of experts, technology scouting, to identify discontinuous technology change (Rohrbeck 2010). For future, it is possible to investigate how the use of social media, patent analysis, and web search engine analysis can be used for idea generation and technology scouting.

Additionally, prioritization of implications of cloud-based digital services from the perspectives of the private sectors such as banks, retailers, and so on to enable the adoption by providing prioritized implications that can be turned into requirements is also a potential area of investigation.